COMMUNICATION TECHNOLOGY, AGE & SOCIAL SUPPORT                              1# The Role of Communication Technology Across the Life Course:

# A Field Guide to Social Support in East York

Anabel Quan-Haase and Molly-Gloria Harper

Sociology Department, Western University

Barry Wellman

NetLab Network, Toronto, Ontario, Canada

Anabel Quan-Haase  https://orcid.org/0000-0002-2560-6709

Molly-Gloria Harper  https://orcid.org/0000-0001-7928-5341

Barry Wellman  https://orcid.org/0000-0001-6340-8837
Quan-Haase, A., Harper, M.-G., & Wellman, B. (in press). The role of communication technology across the life course: A field guide to social support in East York. *Journal of Social and Personal Relationships*.

**Abstract**

We examine how Canadians living in the East York section of Toronto exchange social support. Just as we have had to deconstruct social support to understand its component parts, we now deconstruct how different types of communication technologies play socially supportive roles. We draw on 101 in-depth interviews conducted in 2013-2014 to shed light on the support networks of a sample of East York residents and discern the role of communication technologies in the exchange of different types of social support across age groups. Our findings show that not much has changed since the 1960s in terms of the social ties that our sample of East Yorkers have, and the types of support mobilized via social networks: companionship, small and large services, emotional aid, and financial support. What has changed is how communication technologies interweave in complex ways with different types of social ties (partners, siblings, friends, etc.) to mobilize social support. We found that with siblings and extended kin communication technologies could boost the frequency of interaction and help exchange support at a distance. With friendship ties, communication technologies provide a continuous, constant flow of interaction. We draw implications for theories of social support and for social policy linked to interventions aimed at helping vulnerable groups during the COVID-19 pandemic.

The Role of Communication Technology Across the Life Course:

A Field Guide to Social Support in East York

Exchanging social support in the 21$^{st}$ century is a different enterprise than when our team started studying personal networks in the three earlier East York studies starting in 1968. Although those studies showed variations in the supportiveness of local and physically distant ties (Wellman, 1979), they encompassed only telephone-based support (Wellman & Wortley, 1990), with its limited media richness (Daft & Lengel, 1986). Now exchanging social support in a mediated way is a quasi-given routine part of everyday life and the COVID-19 pandemic has made the reliance even more critical, particularly for vulnerable groups (Robinson et al., 2020). Communication technology with its social affordances – contacting large numbers of friends at once by posting a Facebook status update (Lu & Hampton, 2017) – has facilitated maintaining supportive relations near and far.

Research consistently shows that younger adults are heavy users of communication technology, using many types ranging from social media to texting to video chat (Auxier & Anderson, 2021). However, it remains unclear how adequate communication technologies are for the exchange of different types of social support. Young adults report social media like Facebook cannot serve as a conduit of tangible and esteem support because of its public nature (Liu, Wright, & Hu 2018). This suggests communication technologies are supplemental rather than a replacement for in-person contact and support (Hampton, 2016). Yet, some research suggests that it can serve as a bridge for maintaining a broad network of connections and creating awareness of a need for support (Burrows et al., 2000; Lu & Hampton, 2017). Thus, there is no consensus as to what forms of communication technologies can facilitate or hinder the exchange of what types of social support.

An important omission is that past research has not compared different communication technologies for exchanging a range of support types, as it has mostly zeroed in on one type, often the focus being on Facebook because of its popularity (e.g., Lu & Hampton, 2017). We propose to expand this literature by employing a qualitative approach where participants can report on what types of communication technologies, they use for exchanging different support types rather than limiting the study to a single technology. As not all communication technologies have the same affordances (Fox & McEwan, 2017), we examine how distinct features and functions facilitate or constrain social support exchange mechanisms. To examine and contrast communication technologies more deeply, we draw on the theory of technological affordances (Davis, 2020). Further, Heinze et al. (2015) showed that there are variations in the role of social ties over the life course as sources of support. Social ties are personal connections for the purpose of sharing knowledge, feelings, and experiences (Pescosolido, 2007). This motivates us to also examine how type of tie—e.g., partners, siblings, and friends—mediates the exchange. We address the following research questions:

1. What types of social support do people in different age groups exchange and what role do communication technologies play?
2. What types of social ties use what types of communication technologies for exchanging what types of social support? And, how do technological affordances mediate the exchange?

To answer these questions, we draw on 101 in-person interviews conducted in East York, Toronto, Canada, as part of the fourth wave of East York data collection, which took place in 2013-2014. East York represents a useful case study because it is representative of much of urban Canada, being a multicultural and multiethnic neighborhood. Additionally, research on social relations and the role of communication technology has taken place in East York since the 1960s, thus providing a means for comparative insights. While our past research drawing on the fourth wave of East York data has examined how older adults use communication technology to exchange support (Quan-Haase, Mo, & Wellman, 2017), we have not examined changes over the life span. Other research drawing on the fourth wave of East York data collection has examined mobile use in older adults (Jacobson, Lin, & McEwan, 2017), network structure and composition across the life span (Wellman, Quan-Haase, & Harper, 2020), and digital skills in older adults (Quan-Haase et al., 2018). Yet, it is important to compare age groups because it remains unclear if age differences exist in how various forms of communication technologies are used to exchange different types of support. As best as we know, this is the first exploratory field study reporting on how North American adults use communication technology across age groups to mobilize different types of social support from their personal networks. Our study makes an important contribution in that we study the repertoire of communication technologies used to exchange social support rather than focusing on a single technology. This provides a more holistic understanding of how personal networks are activated for exchanging different types of social support. The present study has important policy implications for how communication technologies are implemented during COVID-19 to aid intergenerational communication and also to provide mediated social support to different age groups.

## Literature Review

**Social Support across Types of Social Ties over the Life Course**

Social support is one of the primary functions of social relations where individuals exchange various types of support (Heaney & Israel, 2008). Much research has pointed toward the benefits of social support (Uchino, 2009; Umberson, Crosnoe, & Reczek, 2010), treating support as either a single concept or seeing it as a variety of resource exchanges (Berkman, Kawachi, & Glymour, 2014). Earlier East York studies contributed to the understanding of social support by proposing a typology of support types, including emotional aid, small and large services, companionship, and financial aid (Wellman, 1979; Wellman & Wortley, 1990). This research showed that the exchange of these types of support depends on the characteristics of a person's network, as different social ties specialize in giving different types and levels of support (Wellman & Wortley, 1990).

Looking closer at the way social support is exchanged and by whom, research has shown that as individuals move throughout the life course, their needs and circumstances change as well as their social roles and relationships (Heinze et al., 2015) and so does the type and amount of social support they give and receive (Kahn & Antonucci, 1980). Though individuals rely on certain sources of support consistently throughout the life course, such as family, friends, neighbors, and community ties (Brajsa et al., 2018), different types of ties take prominence (Heinze et al., 2015). To better understand these differences, we draw from literature focused on middle-aged and older adults. We follow the premises of the technological affordances theory by examining how not only types of communication technologies, but also their specific features and functions mediate the exchange of social support across different age groups.

**Social Support Exchanged via Communication Technologies and Age-related Factors**

When looking at the adoption of communication technologies, there are differences among age groups regarding the exchange of digital forms of support. Young adults (i.e., aged 18 to 35) are more likely to have expansive social networks (Umberson et al., 2010), suggesting that a larger variety of support is available to them. This is because young adults tend to adopt communication technologies to build and maintain their networks (Field, 2008; Umberson et al., 2010). In turn, their larger networks influence the level of perceived support available (Cole et al., 2017). Despite this, studies have suggested young adults do not consider social media to provide tangible and esteem support, rather it provides informational support due to the public nature of some platforms that deter individuals from engaging in more supportive exchanges (Liu et al., 2018). This results in less support exchanged than through in-person interactions (Drouin et al., 2018; Liu & Wei, 2018). For these younger individuals then, the exchange of social support is not necessarily a strong motivation for engaging via communication technologies.

Older adults often face age-related barriers that can cause them to be hesitant in adopting digital devices such as a lack of digital skills, not knowing digital jargon, limited confidence, and the hindrance of small screens and hard to press buttons (Hage et al., 2020; Neves et al., 2019; Quan-Haase et al., 2018). Despite the barriers, studies have found older adults are not a homogenous group. Rather, there is immense variation in terms of digital skills and Internet use (Hargittai & Dobransky, 2017; Quan-Haase et al., 2018). For instance, studies have found receiving support from others, such as "warm experts," helps older adults overcome digital-related fears, feel supported, and gain confidence in using digital technologies (Hanninen, Taipale, & Luostari, 2020), which allows them to take advantage of the many benefits that accompany communication technologies. Even though older adults generally use communication technologies to strengthen existing ties rather than actively seek new ones, evidence shows communication technologies help reduce social isolation, lower levels of depression, enhance support, and promote social inclusion (Choi, Kong, & Jung, 2012), especially for those with limited mobility or in care settings (Burrows et al., 2000; Cotten, Ford, Ford, & Hale, 2014). Even when turning to communication technologies to facilitate the exchange of support, older adults prefer in-person and traditional forms of communication with digital technologies serving as alternative means of connecting with others, particularly when in-person or telephone communication is limited (Baecker, Sellen, Crosskey, Boscart, & Neves, 2014).

While needs may differ throughout life stages, the exchange of support has positive effects on well-being for all age groups because there are more people to communicate with in times of need (Cole et al., 2017; Choi & Noh, 2019). In addition, the exchange of support reduces feelings of loneliness and social isolation by increasing communication and social belongingness among social networks regardless of size (Choi et al., 2012; Choi & Noh, 2019). As a result, perceived support and exchanged support through communication technologies leads to positive outcomes for all age groups. However, little is known about the specific types of communication technologies used to exchange what types of support at different life stages.

## Methods

**East York Context**

The data come from the fourth wave of the East York study. East York is part of the Greater Toronto Area (GTA), home to approximately 6.3 million residents (City of Toronto, 2018), making it one of the largest metropolitan areas in North America. On its own, East York is home to nearly 120,000 residents living in apartment buildings and small houses (Statistics Canada, 2019). Residents of East York reported an average household income of Cdn$113,802, which is greater than Toronto's average annual income of Cdn$102,721 (Toronto City Planning, 2018). East York residents have a median age of 41 years, with 13% of East Yorkers older than age 65 (Statistics Canada, 2019).

**Data Collection**

After obtaining approval from the University of Toronto's Research Ethics Board, we constructed a sampling frame from a random sample of 2,321 East York households provided by Research House, a Toronto-based list-services company. From this list, we randomly contacted 304 people via personal invitation letters sent via mail. From those contacted via follow-up telephone calls or email, 101 agreed to participate in an in-person interview, leading to a 33% response rate. Participants were compensated with a Cdn$50.00 coffee shop gift card for their time.

Qualitative methods were used because social support is perceived (Thoits, 2011); therefore, it is important to understand the perceptions of participants in their own words. In-depth, semi-structured interviews provided flexibility to explore the complex concept of social support via communication technologies (Yeo et al., 2013) because we were interested in learning the full range of communication technologies employed rather than limiting the study to one type. The flexible nature of the interview guide and use of probes allowed interviewers the chance to ask follow-up questions based on participants' responses (Berg & Lune, 2012). Since these were in-depth interviews, each lasted approximately 1.5 hours and contained 65 main questions with additional probes and follow-up questions. A wide range of topics were covered including social networks, technology use, and social support (interview guide available at https://sociodigitaltest.files.wordpress.com/2017/12/interview-schedule_ni-project.pdf). After pilot testing and refining the interview guide, data were collected in 2013-2014. With permission from participants, interviews were recorded and transcribed. To ensure accuracy, one-third of the interviews were checked against the original recording by a trained research assistant. To protect confidentiality of participants, we use pseudonyms reflecting both gender identity and ethnicity.

**Sample**

Our sample comprises 101 English-speaking adults ranging in age from 27 to 93 with a mean and median age of 60 ($SD$ = 15), skewing upwards compared to census data on East York, which report a median age of 41. Table 1 depicts participants' demographic information including gender identity, age, employment status, level of education, enrollment as a student, place of birth, and living arrangements. To examine age-based differences, we divided participants into four groups: under 35 (6 participants), age 35-50 (22 participants), 51-64 years of age (32 participants), and age 65 and older (41 participants). We had an even distribution across men and women and no other gender identity was reported. Reflecting the cultural diversity of Toronto (Toronto City Planning, 2018), participants born outside of Canada were from Europe, the Middle East, Asia, and the Caribbean. In our sample, 96% of participants owned a computer, 95% had a landline, 90% owned a mobile phone, and 37% had a tablet. Ninety-two percent of participants used email, 54% communicated via texting, 46% used video chat like Skype or FaceTime, 57% used Facebook, and 17% used Twitter.

**Table 1. Demographic Characteristics of the Sample (*N*=101).**

| Characteristic | | Number of Participants |
|---|---|---|
| Gender identity | Men | 46 |
| | Women | 55 |
| Age | Under 35 | 6 |
| | 35–50 | 22 |
| | 51–64 | 32 |
| | 65+ | 41 |
| Employment Status | Full-time Work | 42 |
| | Part-time Work | 13 |
| | Unemployed | 8 |
| | Retired | 37 |
| Level of Education | High School | 22 |
| | College | 17 |
| | University | 40 |
| | Postgraduate education | 22 |
| Current Student | Yes | 12 |
| | No | 89 |
| Place of Birth | Ontario | 45 |
| | Canada, Outside Ontario | 10 |
| | USA | 7 |
| | Outside of Canada or USA | 37 |
| Living Arrangement | Alone, never married or divorced | 29 |
| | Lives with partner/spouse | 32 |
| | Lives with partner and child(ren) | 25 |
| | Alternative living arrangement (e.g., alone with children, siblings, relatives) | 15 |

**Data Analysis**

To stay close to the data, our analysis was guided by thematic analysis (Braun & Clarke, 2006). Research assistants met regularly over two years, discussed any coding discrepancies, and refined any codes when necessary to converge on a final list of codes. In the final stage of coding, all team members reviewed the final list of themes for coherence and provided guidance in selecting meaningful quotes that highlighted key findings (Braun & Clarke, 2006). To enhance the trustworthiness of the data, we used thick description of participants' contexts and everyday life circumstances (Houghton, Casey, Shaw, & Murphy, 2013).

## Communication Technologies and Social Support

**RQ1: What types of social support do people in different age groups exchange and what role do communication technologies play?**

*The Prevalence of Different Types of Support*

Participants predominantly exchanged three types of social support: companionship, small services, and emotional aid. Large services and financial aid were exchanged to a much lesser extent. Remarkably, these types of support, as well as the individuals who exchanged them, has been roughly consistent in all East York studies, starting with evidence collected in 1968 (Wellman & Wortley, 1990). We did, however, notice slight discrepancies between reported giving and receiving, with participants overall reporting more giving than receiving for companionship, small services, emotional aid, and financial aid (see Table 2).

**Table 2. Types of Social Support Exchanged (Giving and Receiving) Across Age Groups**

Types of Social Support (%)

| Age groups | Companionship | Small services | Large services | Emotional aid | Financial aid |
|---|---|---|---|---|---|
| **Under 35** (*n*=6) | | | | | |
| Giving | 100 | 67 | 17 | 67 | 67 |
| Receiving | 100 | 83 | 17 | 67 | 33 |
| **35–50** (*n*=22) | | | | | |
| Giving | 95 | 68 | 36 | 36 | 18 |
| Receiving | 91 | 64 | 32 | 32 | 9 |
| **51–64** (*n*=32) | | | | | |
| Giving | 75 | 75 | 22 | 56 | 13 |
| Receiving | 72 | 75 | 34 | 28 | 0 |
| **65+** (*n*=41) | | | | | |
| Giving | 52 | 76 | 73 | 50 | 2 |
| Receiving | 60 | 33 | 40 | 50 | 0 |
| **Overall** (*N*=101) | | | | | |
| Giving | 66 | 57 | 18 | 39 | 12 |
| Receiving | 63 | 50 | 20 | 24 | 4 |
| Examples | -Visiting/being visited -Attending parties or occasions -Organize get-togethers | -Babysitting children/pets -Advice -Household chores -Sharing baked goods | -Caring for ill friends and family -Offering a place to stay -Renovating -Childcare | -Support during difficult times such as bereavement, illness, and conflicts | -Lending or borrowing money to/from family and/or friends for various large purchases |

*Companionship*

Across all age groups, participants exchanged ***companionship*** reciprocally at a much greater extent than all other types of support. Participants stressed that having companionship through communication technologies allowed them to strengthen relationships with all social ties, but in particular with relatives that would otherwise only be seen annually at family functions. Participants were able to maintain ties with relatives even when circumstances like geographical distance limited face-to-face contact. For instance, Meike Hallberg (P48, W, 53) stated that video chats via FaceTime worked better than phoning to bring her and her husband together with other family members outside of Toronto. The visual features of FaceTime allowed Meike to hear and see her brother and sister-in-law on a single screen, which afforded a sense of co-presence in comparison to the phone. Another important feature of FaceTime is that it facilitated not only dyadic communication, but also digital group video chat. In that way, a key affordance is that it can expand dyadic communication, allowing larger groups to partake in interaction and providing a sense of group companionship:

> *What is nice about FaceTime is sometimes [my husband] will sit next to me or [my sister-in-law] will sit next to [my brother] so then the four of us chat. So that's nice. Which we wouldn't do on the phone.*

We found that since a majority of those aged 35 to 50 lived in the same household as their partners and often children, face-to-face was preferred over communication technologies for companionship. By contrast, their friends often lived beyond their neighborhoods throughout the metropolitan area, the province of Ontario, and internationally. This is in stark contrast to earlier East York data (Wellman, 1979), where distant ties were more likely to be with kin rather than with friends. To stay in touch with friends throughout the metropolitan area of Toronto, participants used multiple communication technologies to exchange companionship and to organize in-person get-togethers such as dinners and parties.

For companionship, participants used communication technologies primarily for socializing with existing ties rather than with new ones, and this was consistent across all age groups. We found that exchanging companionship through digital technologies was more often with friends than with close and extended kin, with whom companionship was exchanged more frequently in-person or via phones. Yet with age, companionship declined (Table 2). Those aged 65+ reported the lowest levels of companionship (52% and 60% for giving and receiving, respectively) in comparison to all other age groups. For example, those under age 35 reported 100% for giving and receiving. Older adult participants reported several barriers to companionship including an impaired ability to travel themselves or their similar-aged kith and kin, and the residential dispersion after retirement of themselves, their adult children, and their peers. These barriers made communication technologies particularly relevant for this age group, both for directly exchanging companionship, but also to coordinate visits.

### Small Services

Many participants exchanged small services—the second most prevalent form of support. In contrast to companionship, communication technologies scarcely played a role in this form of social support. Rather, exchanging small services was characteristic of participants' face-to-face neighborly ties and consisted of household jobs such as helping with yard work, minor house repairs, and babysitting children and/or pets. These were reciprocal exchanges, with many participants offering help to their neighbors when it was needed because they could count on similar help in return. Those 65+ were more likely to give small services than receive—mostly in-person, which reflected their willingness to help family and friends, as they felt they had the time to engage in altruistic acts. Despite geographic closeness, for some participants phone, email, and texting added onto face-to-face for neighborly support exchange.

*Emotional Aid*

Emotional aid was both widely given and received by most participants and was the third most prevalent form of support. In addition to face-to-face contact, participants used a variety of communication technologies to for exchanging emotional aid, especially when they needed or offered an ear or shoulder during difficult times such as illnesses, bereavements, or major troubles. Following in-person encounters, calling--both mobile and landline--were perceived as beneficial, even if not preferred, for exchanging emotional aid because people felt there was more of a connection when they could hear someone's voice and/or see someone's facial expressions. These social affordances facilitated closeness and intimacy, which for participants allowed emotions to be expressed. Many did not see texting or emailing as conducive for providing emotional aid because of the lack of intonation and nonverbal cues. This suggests that social affordances of co-presence provided auditorily or visually through calling or video chatting are critical for the provision of emotional aid.

> *Well, with a text message, it can be a little, sometimes, you can't read the emotion into texts. And you can't read emotion into email. But if you're talking to somebody on the phone, you can sense their emotions and you can tell if there's a problem or you can tell if they're happy (Maggie Darling, P22, W, 60)*

*Large Services*

The much rarer exchange of large services pertained more to family matters, such as availability to care for elderly parents or kin, babysitting children, and offering a place to stay when out-of-town guests visited. The delivery of large services occurred in-person, but many used communication technologies to create awareness about the needed help and to organize the exchange. This finding parallels past studies that demonstrated how Facebook is a valuable tool for disclosing a need for help (Lu & Hampton, 2017), but in-person continues to be the means of exchanging large services. For example, Maggie Bethany (P75, W, 55) is close with her mother and visits her frequently to provide help.

> *When it's my family, I see my mother fairly often cause she's ninety and she needs some help…My mother would have me there every day.*

Communication technologies, such as mobile phones, were tools for facilitating social accessibility and reachability for both emergency and more routine large services. Participants reported worrying about their parents and children being able to reach them anywhere anytime should a serious situation arise. Thus, communication technologies serve as an important intergenerational bridge that affords constant connectivity via texting or cell phone. For instance, Dan Pouchik (P64, M, 54) replaced his landline with a mobile phone so he could be always reachable.

> *The reason I would get a cell phone is to replace my landline…I have an elderly mother [and] if I have to be in Hamilton, at a moment's notice, and they need to contact me, then, my phone is portable.*

**Financial Aid**

Financial aid was limited among participants and often coordinated via technologies including web sites. Those under 35 were the most likely to give and receive financial support from family, friends, or relatives. Across all age groups, lending and borrowing of money often took place for large types of purchases, and for sending financial aid to geographically distant, often overseas, family members. Those few who gave financial support often did so without question, implying it is part of their familial commitment or part of being a friend.

**RQ2: What types of social ties use what types of communication technologies for exchanging what types of social support? And, how do technological affordances mediate the exchange?**

*Children and Spouses/Partners*

Participants exchanged social support with their children (under age 18) and their partners/spouses who lived in the same household primarily in-person and secondarily through one-on-one phone calls and texting. For example, Dan Pouchik (P64, M, 54) texted his ex-wife for coordinating pick-up/drop-off times for their children. Communication technologies were often used to coordinate activities and the exchange of small services rather than to provide companionship or emotional aid. Other forms of communication technologies, however, substituted companionship when in-person contact was disrupted. Henry Macdonald (P9, M, 42) said when he is away for work, his preference for keeping in touch with family is Skype because of the co-presence afforded:

> *When I'm travelling then I would be more inclined to use Skype to call my family and call my partner.*

As children grew up and moved out of the family home, communication technology was used to keep in touch for companionship and emotional aid. For instance, participants such as Michael Harris (P4, M, 56) and Maggie Bethany (P75, W, 55) said they frequently texted and called their children (aged 18 and older) to talk while they were away at university. Participants also complemented phone calls and texts with video chats, some email, and limited Facebook engagement. This suggests that these strong ties were connected through multiple media providing evidence of media multiplexity. What communication technology was preferred depended on the circumstances, urgency, and whether the exchange of support was being coordinated or actual support was being exchanged. Except for Facebook, these were mostly private dyadic interactions. Participants could contact their children to exchange companionship (e.g., when they missed them) or emotional aid by integrating various forms of communication technologies.

*Parents*

Participants reported that in-person contact with their parents declined as physical distance increased. Communication mostly occurred via telephone, which was often attributed to their parents' lack of skill with communication technologies. For example, Patricia Long (P18, W, 33) relied on the phone to call her mother because this was the only way to contact her.

> *Well with my mom, she's not really into technology...She refuses to get a cell phone [saying] 'no, I don't need one, I'm sixty years old.' She just refuses so it's [telephone] the only way I can contact her.*

Even though participants' parents were more likely to use landline phones to call them, participants themselves used their mobile phones to be as available as possible to their parents. Some actively encouraged their parents to adopt mobile phones, or even got mobile phones for them, to increase their mutual accessibility should an emergency arise or if there was an immediate need for support. In this way, communication technologies afforded instant communication and peace of mind.

> *My parents, for example, I just bought them a cell phone off of my employee home mobile. So, I just switched their home phone to a cell phone just in case we were to go, and I could have contact with them (Abbas Farrukh, P17, M, 27)*

For about half of study participants, phone calling was not all that was used to connect with their (often elderly) parents, rather they showed media multiplexity, that is, they connected with their parents via text, email, video chat, and social media (Facebook and instant messaging), which made communication easier with geographically distant ties, those outside the Greater Toronto Area (GTA).

*Grandchildren*

Study participants reported that the exchange of social support with grandchildren was important and used communication technologies as a mediator. For example, Andor Millos (P71, M, 64) said that he generally only sees his family occasionally, but using FaceTime allowed him and his wife to see their children and grandchildren in-between in-person contact. The immediacy of video chat afforded a sense of co-presence, which is important when exchanging companionship.

> *My wife is thrilled that we can use FaceTime with [the iPad]. We just became grandparents within the last six months, so my wife sees the baby all the time on the iPad.*

By using communication technologies, parents could communicate with their children and grandchildren through email, video chat, and social media. For instance, participants such as Mary Orbison (P40, W, 67), enjoyed seeing family photos on Facebook because it made them feel closer and more involved.

> *I like looking at pictures of them and their kids and their kids' kids.*

We see that communication technologies were used in addition to or in place of phone calling, which led to greater tie closeness by increasing the overall frequency and quality of communication. Both increased frequency of communication and closeness are core elements of tie closeness (Marsden & Campbell, 1984). In particular, activities such as video chatting and looking at pictures on Facebook led to feelings of closeness thereby strengthening relationships. Taken together, the sections on parents and grandchildren show the importance of communication technologies for intergenerational communication.

***Siblings and Extended Kin***

In one's network, there are usually more siblings and extended kin (e.g., cousins, aunts/uncles, grandparents) than parents and children, but these ties are often widely dispersed rather than in the same neighborhood (Mok, Wellman, & Carrasco, 2010). Communication technology helped break down geographical barriers to communication, which was often restricted to family gatherings like holidays, celebrations, and vacations. Phone calling with relatives at a distance was perceived as much easier and cheaper than in the past because of low cost long distance plans. This has increased the volume of communication with kin such as Duncan Robertson's (P33, M, 83) contact with family in England.

> *Cheap telephone really makes things so much better, but now, if you know what to do, it's much cheaper.*

Other forms of communication technology such as video chat were free, making them a go-to for contact with non-local relatives. Middle-aged and older adults preferred traditional phone calling over email and considered the phone key for emotional support. Participants stressed the limited affordances of email in comparison to the phone as a detriment for exchanging emotional support: few cues, no voice intonation, and in particular its asynchronous nature. For example, Catherine O'Henly (P53, W, 67) preferred phoning because she likes to talk directly to her siblings:

> *I like to talk with them. If you send someone an email, you're not talking with them. They'll get it and then they read it and respond. I prefer to call up on the phone and talk to them directly.*

For some older adults, however, email's social affordances were valued. As their networks were geographically dispersed, they valued email's ease of use, its low cost, and ability to contact multiple relatives at once.

> *Because it's easier and it's cheaper and you can email multiple members (Benjamin Jones, P31, M, 80)*

Older adults also often discussed email's low intrusion as an advantage. The sense of co-presence in the exchange of companionship was another major affordance of communication technologies for keeping up with siblings and extended kin, reported across all age groups. Participants often saw video chat as being the next best thing to physically being there because it afforded participants to see and hear their relatives on the other end compared to just voices on the phone. For example, Aaron Collins (P3, M, 69), who has a brother and a sister who he sees in-person only once a year, told us that he preferred the audiovisual media richness and emotional cues afforded by Skype video chat compared to email which is text-based:

> *I like Skype-like communication because you can tell a lot about the other person by watching. You can tell if they're nervous, or unhappy, or whatever.*

**Friends**

East York is not an urban village, rather most participants had friends dispersed through the Greater Toronto Area, Canada, and other parts of the world. While all age groups preferred face-to-face contact, and used it with those nearby, communication technologies complimented their connectivity and made maintaining friendship connections, keeping in touch, and exchanging support-- especially emotional support--much easier. It is important to specify, however, that these were usually existing friendships as very few used communication technologies to find new ones online -- when they did, it was primarily those who were dating online who wanted to meet new people. Thus, time spent online was a prominent means for friends to engage, communicate, identify needs, and coordinate plans. For instance, Maggie Bethany (P75, W, 55) said the computer was her most important device because it allowed her to communicate.

> *That's how I communicate with people. I enjoy Facebook. I don't post a lot of things on Facebook, but I like to hear what my friends are doing, and I really connect with people that way.*

Using phone calls, emails, texts, and Facebook as the primary tools used to connect with friends, made planning easier and smoother due to the social affordances that communication technologies provide, such as multi-group emailing and messaging, quickness of texting, and the ability to contact a friend on their mobile phone for last minute plans. For example, Duncan Robertson (P33, M, 83) said that emailing enabled him to stay in touch easily:

> *I can keep in touch with friends much easier and it's funny—some of my friends have funny stories and they send them, so they just forward them to me…It's a way of keeping in touch…So people who I wouldn't normally communicate with—the contact is enhanced by forwarding these pictures or these jokes or stories. So that helps a lot.*

Across age groups, participants used phone calling most often when communicating with friends due to its immediacy and nuance of responses. However, there are some differences worth noting, specifically among younger and older generations. Younger participants expressed their desire to text, valuing its quick and unobtrusive nature, feeling it to be equivalent to phonecalling. They liked the affordance of keeping a communication stream going while engaging in other activities. By contrast, older generations (65+) did not rely on texting (except for 1 participant) because they felt it was not as emotionally fulfilling as a phone call or in-person conversation, for them quality mattered over frequency:

> *I like to hear the person talk and have a real conversation, and to me, if you're doing it with texting, it's still not the same as a real conversation… (Olga Kurt, P37, W, 66)*

Facebook was used for basic communication with friends online such as exchanging information, group messaging, planning of events, or commenting on photos and updates. As well, it was commonly used for exchanging social support with physically distant friends. Yet, people often preferred phoning over Facebook because they could reach someone right away, which they saw as important when needing emotional support. While Facebook was perceived to be a suitable alternative to texting and phone calling, participants like Trudy Wright (P89, W, 57) told us, Facebook's purpose was more for accessing information rather than exchanging companionship or emotional support because it does not allow for in-depth conversations.

> *I don't post a lot of things on Facebook, so I don't feel it's a vehicle for communication; it's a vehicle for information…For me it doesn't foster a two-way street of communication.*

Despite mixed attitudes, we found two-thirds of the younger and middle-aged participants recognized that Facebook provides opportunities for increasing engagement with social networks and informing friends of their need for support, even if it was not used explicitly for exchanging social support.

For younger participants, networking across groups of friends and social settings was a key affordance of communication technologies. For instance, participants could be Facebook chatting with a group of friends while talking on the phone with another friend, or they could be in-person with their friend(s) and texting other groups of friends simultaneously.

*Neighbors*

Geography and proximity were key in the choices participants made when communicating across their diverse array of social networks, appearing to be the factor that determined the use of communication technologies to maintain connections. This was highlighted in participants' attitudes toward associating with their neighbors. The exchange of support with neighbors was heavily face-to-face, with only 14 participants—primarily those 51+—indicating they turned to technologies for communicating with neighbors, which was a way of transcending mobility impairment. They usually felt it unnecessary to share contact information with neighbors because of physical closeness. For instance, David Hawthorne (P74, M, 61) says he does not have any neighbors' phone numbers and only communicates with his neighbors when he sees them outside. They did this knowing that they could turn to their neighbors should support (primarily small services) be needed.

However, there was an exception for middle-age and older adults who benefitted from the affordances that communication technologies provide for neighborly ties. For these two age groups, the distinction between neighbor and friend was blurred, relying on their neighbors for companionship. In such cases, phone calling, and texting, allowed for check-in calls, facilitating in-person contact, and organizing of small exchanges of support. For example, Mariam Roth (P94, W, 54) and Trudy Wright (P89, F, 57) both have close friends living in their immediate neighborhoods, using a quick text or mobile phone call to confirm plans to see each other.

*Workmates*

For co-workers, participants used a combination of phone calling, texting, emails, and face-to-face to communicate. Aside from in-person contact while at work or attending work-related social events, participants deemed email the primary and most convenient method of communication with colleagues because of its asynchronicity—the ability to pass along important messages without disturbing one another.

## Discussion

We investigated exchanging diverse types of social support in the 21$^{st}$ century by examining the role of communication technologies across age groups as well as how type of social tie mediates the exchange. Like earlier East York studies (Wellman & Wortley, 1990), we found the same types of support (i.e., companionship, emotional aid, and small services) are relevant as well as the individuals who exchange them. We did notice slight discrepancies between reported giving and receiving, with participants overall reporting more giving than receiving for companionship, small services, emotional aid, and financial aid. There is, however, one notable exception. We found older adults to stand out whereby their giving of small and large services exceeded their receiving. For many older adults, giving support was an important part of their everyday lives, regardless of social tie (i.e., family ties, neighbors). In some ways, they were making deposits into the bank of support, with the hope and expectation that they would receive it in return when needed (Quan-Haase et. al, 2017). We did find in agreement with past studies that older adults received companionship from their family, friends, and neighbors (Bromell & Cagney, 2014).

Despite Marshall McLuhan's oracular pronouncement, we found "the medium is [not] the message" (1964, 12). Rather, the various communication technologies partially shape, facilitate, and hinder the conveyance of various types of support. Communication technologies do not provide support; ties provide support. But communication technology is often the delivery system through which support flows, and is even more often the infrastructure that maintains the ties, both strong and weak, that can be mobilized for support (Hampton, 2016; Liu et al., 2018). With the partial exception of gamers (Nardi, 2010), there is no reason for social scientists to study technology-mediated social networks as isolated phenomena. It is the relationships that are important, and supportive ties integrate a range of technologies for connectivity with face-to-face encounters.

Our findings reveal that face-to-face contact is the preferred means for support exchange across all age groups. Yet, we also find that participants are adopting alternative digital ways for a range of reasons, such as how communication technologies lessen geographical barriers, which affords accessibility, reachability, and facilitates more frequent communication across their social connections (Hampton et al., 2011; Liu & Wei, 2018). Participants kept telling us, communication technologies have made supportive communication easier, more frequent, and increasingly convenient--"cause it's easier" (Trudy Wright P89, W, 57). The increased frequency has made social ties more resilient than in the past where moving homes, changing jobs, and other life changes could disrupt relations (Brajsa et al., 2018; Cornwell & Goldman, 2020; Rozzell et al., 2014), which is especially true amidst the COVID-19 pandemic (Robinson et al., 2020).

Even though the proportion of ties exchanging each type of support has remained roughly consistent (see Wellman & Wortley, 1990), we found participants used communication technologies to strengthen connectivity mostly with their existing ties, predominantly those with whom they have stronger connections. However, communication technologies were not equal in their adoption for communication and supportive exchanges. For example, for some social ties, like siblings and extended kin, in-person gatherings were restricted to annual events, meaning in-person communication was severely limited, but communication technologies could boost interactions to happen more frequently and help exchange support at a distance. In this way, communication technologies work well, allowing individuals to coordinate exchanges of support through a variety of digital outlets. For other social connections, like friendship ties, there were different dynamics. Rather than relying on communication technologies to fill-in the time between in-person, East Yorkers readily integrated a variety of communication technologies for a continuous, constant flow of communication including both quick, rapid exchanges as well as more intimate exchanges of support. Navigating both public and private communication channels allowed individuals to seamlessly maintain these social connections without any real disruption in communication.

Past studies have focused extensively on Facebook, showing that Facebook increases the perceived amount of support that is available via one's social network (Lu & Hampton, 2017). Our findings support and expand these findings showing that across all age groups Facebook helped exchange support: young adults used it mostly with friends, while middle-aged adults (35-64) used it with children, extended kin, and friends, and older adults—only a few—used it with children, extended kin, and friends. Facebook when examined in relation to other communication technologies, however, only plays a limited role. In fact, East Yorkers across all age groups turn to and prioritize more personal, dyadic communication technologies (e.g., phone and video chat) with their stronger ties like children, partners, siblings, parents, and close friends. This is because despite the many informational benefits Facebook provides, its asynchronicity and broadcast nature (e.g., lack of privacy) hinder the exchange of companionship and emotional aid.

Even though communication technologies like Facebook expand one's reach into networks of support, Cornwell and Goldman (2020) found local ties are uniquely positioned to provide companionship. While our study supports Cornwell and Goldman's (2020) findings, we also found that for middle-aged adults and older adults communication technologies enhance local ties. This supports findings from Netville (Hampton & Wellman, 2003), a wired neighborhood, suggesting that communication technologies are not radically transforming communities, nor replacing local, neighborly ties, but rather are adding to in-person communication. East Yorkers use these technologies to reap many benefits from their local ties, including exchanging support, even if used to arrange for in-person support.

Social support studies suggest that the quality of relations is critical for individuals' well-being rather than the sheer volume of interaction (Pinquart & Sörensen, 2001). Our findings suggest that communication technologies with their affordances of co-presence contribute to the enhancement of relations, whereas communication technologies that afford speed and convenience mostly contribute to quantity but not quality. But this finding is more nuanced. While for all age groups co-presence matters, older adults give this even more prominence when it comes to exchanges of companionship and emotional aid. By contrast, those aged 35 to 64 rely on multiple communication technologies to exchange all types of social support, choosing the type depending on social context and current need. Also, younger age groups see the benefits of frequent exchanges for increasing the resilience of social ties. In that way, technologies and their affordances do not dictate how support is exchanged for younger age groups, rather all types are used flexibly to exchange support with close ties. This suggests that age-related factors like familiarity with technologies and established habits structure how technologies are integrated into support networks.

As is almost always the case, our study has its limitation. The data were collected in 2013-2014, and thus future research needs to expand the scope by integrating newer technologies like TikTok and also documenting the social impacts of the pandemic. In our interviews, we do not have complete data for participants' race/ethnicity, sexual orientation, and disability information, which are all dimensions of importance for the study of social support and technology use/adoption (Robinson, et al., 2020). Further, our case study is limited to Toronto, a metropolitan area that is unique in its multicultural composition and high rate of recent immigrants/refugees, and thus research needs to be expanded to other locales.

On the positive side, through 101 interviews with East York residents, our case study provided rich data on aspects of the pressing social support questions that the COVID-19 pandemic has further brought to the forefront. We are proud to have followed Bayer, Triệu, and Ellison (2020) by pursuing theoretically-driven research approaches with the aim "to produce enduring knowledge about social technologies" (p. 472).

**Acknowledgement**

This paper draws on research supported by the Social Sciences and Humanities Research Council of Canada. We highly value the contributions of our interviewers, transcribers, and coders. We are grateful to the East Yorkers who took the time to share their experiences with our team.